\documentstyle[prc,aps]{revtex}
\draft

\newcommand{\ba}{\begin{eqnarray}}
\newcommand{\ea}{\end{eqnarray}}
\input{psfig}

\begin{document}

\title{Collective states in nuclei and many-body 
random interactions}

\author{R. Bijker$^1$ and A. Frank$^{1,2}$} 

\address{$^1$Instituto de Ciencias Nucleares, 
Universidad Nacional Aut\'onoma de M\'exico, \\
Apartado Postal 70-543, 04510 M\'exico, D.F., M\'exico \\ 
$^2$Centro de Ciencias F{\'{\i}}sicas, 
Universidad Nacional Aut\'onoma de M\'exico, \\
Apartado Postal 139-B, Cuernavaca, Morelos, M\'exico}

\date{April 3, 2000}

\maketitle

\begin{abstract}
Low-lying collective states in nuclei are investigated in the 
framework of the interacting boson model using an ensemble of 
random many-body interactions. It is shown that whenever the 
number of bosons is sufficiently large compared to the rank 
of the interactions, the spectral properties are characterized 
by a dominance of $L^P=0^+$ ground states and the occurrence 
of both vibrational and rotational band structures. This indicates 
that these features represent a general and robust property of the 
collective model space. 
\end{abstract} 


\section{Introduction}

Random matrix ensembles provide a powerful tool to study generic 
spectral properties of complex many-body systems \cite{review}. 
Most applications in the literature have centered on global 
characteristics such as first neighbor energy distributions, 
which typically involve states with 
the same quantum numbers (angular momentum, parity, isospin, ...). 
Recently, the relation between low-lying states in even-even nuclei 
with different quantum numbers was examined using Hamiltonians 
with random interactions in the nuclear shell model (SM) 
\cite{JBD,CJ,JBDT,BFP2} and the interacting boson model (IBM) 
\cite{BF,oax00}. These studies  have given rise to several surprising 
results. In both cases it was found that for a large variety of 
conditions there is a dominance ($\stackrel{>}{\sim} 60 \%$) 
of $L^P=0^+$ ground states despite the random nature 
of the interactions. In addition, in the 
SM strong evidence was found for the occurrence of pairing 
properties \cite{JBDT}, and in the IBM for both vibrational and 
rotational band structures \cite{BF}. These results are not only 
based on energies, but also involve the behavior of the wave 
functions via the 
pair transfer amplitudes in the case of pairing, and the quadrupole 
transitions for the collective bands. The use of random interactions 
(both in size and in sign) show that these regular features arise 
for a much wider class of Hamiltonians than are usually 
considered to be realisitic. 
These results are in qualitative agreement with the empirical 
observations of very robust features in the low-lying spectra of 
medium and heavy even-even nuclei and a tripartite classification 
in terms of a seniority, a vibrator and a rotor regime 
\cite{Casten,Zamfir}. 
The conventional wisdom in nuclear structure physics is that the 
observed properties of nuclei can be explained by specific 
features of the SM (or IBM) Hamiltonian. The studies with random 
interactions, however, seem to imply that some of the generic 
characteristics of these systems may already be encoded in the 
corresponding shell model (or $sd$ boson model) space. 
This is particularly striking in the case of the IBM, for which 
the model space corresponds to a drastic truncation of the original 
(shell model) Hilbert space to that composed of like-nucleon pairs 
with angular momentum $L=0$ and $L=2$ \cite{IBM,Talmi}. 
The selection of such a restricted subspace seems to impose strong 
constraints on the possible spectral properties.  

These considerations lead naturally to the question of what are the 
specific causes of this behavior, given that the ingredients of the 
calculations are the preservation of fundamental symmetries of the 
Hamiltonian (hermiticity, rotational invariance, time-reversal 
invariance), the one- and two-body nature of the Hamiltonian, a 
given number of active particles, and the structure of 
the model space. In \cite{BFP2} it was shown that the preponderance 
of $L^P=0^+$ ground states in the nuclear shell model is 
not due to the time-reversal symmetry of the interactions. 
The purpose of this paper is to address explicitly the role of the 
particle number and the rank of the random many-body 
interactions on the systematics of collective states in nuclei. 

\section{Random interactions in the IBM}

To study the global features of low-lying collective states 
in nuclei we carry out an analysis of the IBM with random 
interactions. In the IBM, collective nuclei are described as a 
system of $N$ interacting monopole and quadrupole bosons \cite{IBM}. 
We consider all possible one-, two- and three-body interactions. 
The one-body Hamiltonian contains the boson energies 
\ba 
H_1 &=& \epsilon_0 \, s^{\dagger} s  
+ \epsilon_2 \sum_m d^{\dagger}_m d_m ~. 
\label{h1}
\ea
The two-body interactions can be expressed as 
\ba
H_2 &=& \sum_{L=0,2,4} \, \sum_{i \leq j} \zeta_{L_{ij}} \, 
\frac{P^{\dagger}_{L_i} \cdot \tilde{P}_{L_j} 
+ P^{\dagger}_{L_j} \cdot \tilde{P}_{L_i}}{1+\delta_{ij}} ~, 
\label{h2}
\ea
with $\tilde{P}_{LM}=(-1)^{L-M} P_{L,-M}$. Here  
$P^{\dagger}_L$ denotes the creation operator of a pair of bosons 
coupled to angular momentum $L$ 
\ba
P^{\dagger}_{0_1} &=& \frac{1}{\sqrt{2}} \,  
(s^{\dagger} \times s^{\dagger})^{(0)} ~, 
\nonumber\\
P^{\dagger}_{0_2} &=& \frac{1}{\sqrt{2}} \, 
(d^{\dagger} \times d^{\dagger})^{(0)} ~, 
\nonumber\\
P^{\dagger}_{2_1} &=& 
(s^{\dagger} \times d^{\dagger})^{(2)} ~, 
\nonumber\\
P^{\dagger}_{2_2} &=& \frac{1}{\sqrt{2}} \, 
(d^{\dagger} \times d^{\dagger})^{(2)} ~, 
\nonumber\\
P^{\dagger}_{4_1} &=& \frac{1}{\sqrt{2}} \, 
(d^{\dagger} \times d^{\dagger})^{(4)} ~.
\ea
Similarly, the three-body interactions are given by
\ba
H_3 &=& \sum_{L=0,2,3,4,6} \, \sum_{i \leq j} \xi_{L_{ij}} \, 
\frac{P^{\dagger}_{L_i} \cdot \tilde{P}_{L_j} 
+ P^{\dagger}_{L_j} \cdot \tilde{P}_{L_i}}{1+\delta_{ij}} ~, 
\label{h3}
\ea
with
\ba
P^{\dagger}_{0_1} &=& \frac{1}{\sqrt{6}} \, 
(s^{\dagger} \times s^{\dagger} \times s^{\dagger})^{(0)} ~, 
\nonumber\\
P^{\dagger}_{0_2} &=& \frac{1}{\sqrt{2}} \, 
(s^{\dagger} \times d^{\dagger} \times d^{\dagger})^{(0)} ~, 
\nonumber\\
P^{\dagger}_{0_3} &=& \frac{1}{\sqrt{6}} \, 
(d^{\dagger} \times d^{\dagger} \times d^{\dagger})^{(0)} ~, 
\nonumber\\
P^{\dagger}_{2_1} &=& \frac{1}{\sqrt{2}} \, 
(s^{\dagger} \times s^{\dagger} \times d^{\dagger})^{(2)} ~, 
\nonumber\\
P^{\dagger}_{2_2} &=& \frac{1}{\sqrt{2}} \, 
(s^{\dagger} \times d^{\dagger} \times d^{\dagger})^{(2)} ~, 
\nonumber\\
P^{\dagger}_{2_3} &=& \frac{1}{\sqrt{6}} \, 
(d^{\dagger} \times d^{\dagger} \times d^{\dagger})^{(2)} ~, 
\nonumber\\
P^{\dagger}_{3_1} &=& \frac{1}{\sqrt{6}} \, 
(d^{\dagger} \times d^{\dagger} \times d^{\dagger})^{(3)} ~, 
\nonumber\\
P^{\dagger}_{4_1} &=& \frac{1}{\sqrt{2}} \, 
(s^{\dagger} \times d^{\dagger} \times d^{\dagger})^{(4)} ~, 
\nonumber\\
P^{\dagger}_{4_2} &=& \frac{1}{\sqrt{6}} \, 
(d^{\dagger} \times d^{\dagger} \times d^{\dagger})^{(4)} ~, 
\nonumber\\
P^{\dagger}_{6_1} &=& \frac{1}{\sqrt{6}} \, 
(d^{\dagger} \times d^{\dagger} \times d^{\dagger})^{(6)} ~. 
\ea
The coefficients $\epsilon_L$, $\zeta_{L_{ij}}$ and $\xi_{L_{ij}}$ 
correspond to the 2 one-body, 7 two-body and 17 three-body 
matrix elements, respectively. They are chosen independently from 
a Gaussian distribution of random numbers  
with zero mean and variance $v^2$ as 
\ba
\left< \epsilon_L \epsilon_{L'} \right> &=& 
\delta_{LL'} \, 2 \, v^2 ~, 
\nonumber\\
\left< \zeta_{L_{ij}} \zeta_{L'_{i'j'}} \right> &=& 
\delta_{LL'} \, (1+\delta_{ij,i'j'}) \, v^2 ~, 
\nonumber\\
\left< \xi_{L_{ij}} \xi_{L'_{i'j'}} \right> &=& 
\delta_{LL'} \, (1+\delta_{ij,i'j'}) \, v^2 ~,
\nonumber\\
\left< \epsilon_{L'} \zeta_{L_{ij}} \right> &=& 
\left< \epsilon_{L'} \xi_{L_{ij}} \right> \;=\; 
\left< \zeta_{L_{ij}} \xi_{L'_{i'j'}} \right> \;=\; 0 ~.
\label{ensemble}
\ea
The choice of the ensembles is such that they are invariant under 
orthogonal basis transformations. The variance of the Gaussian 
distribution $v^2$ sets the overall energy scale. 
The ensemble defined by Eq.~(\ref{ensemble}) for $H_k$ 
is called the $k$-body random ensemble ($k$-BRE) \cite{French}.  
For two-body interactions $H=H_2$ it reduces to the T(wo)BRE   
\cite{French,Bohigas}. When the number of bosons is equal to the 
rank of the interactions $N=k$, the Hamiltonian matrix is entirely 
random and the ensemble coincides with the Gaussian orthogonal 
ensemble (GOE). For $N>k$ the many-body matrix elements of 
$H_k$ are correlated via the appropriate reduction formulas and 
depend, in principle, on all random $k$-body matrix elements. 

\section{Results}

In \cite{BF} we used random one- and two-body interactions with 
$N=16$ to study the systematics of low-lying collective states 
in the IBM. 
Here we wish to study how these results depend on the boson number 
and the rank of the random interactions. 

\begin{figure}
\centerline{\hbox{
\psfig{figure=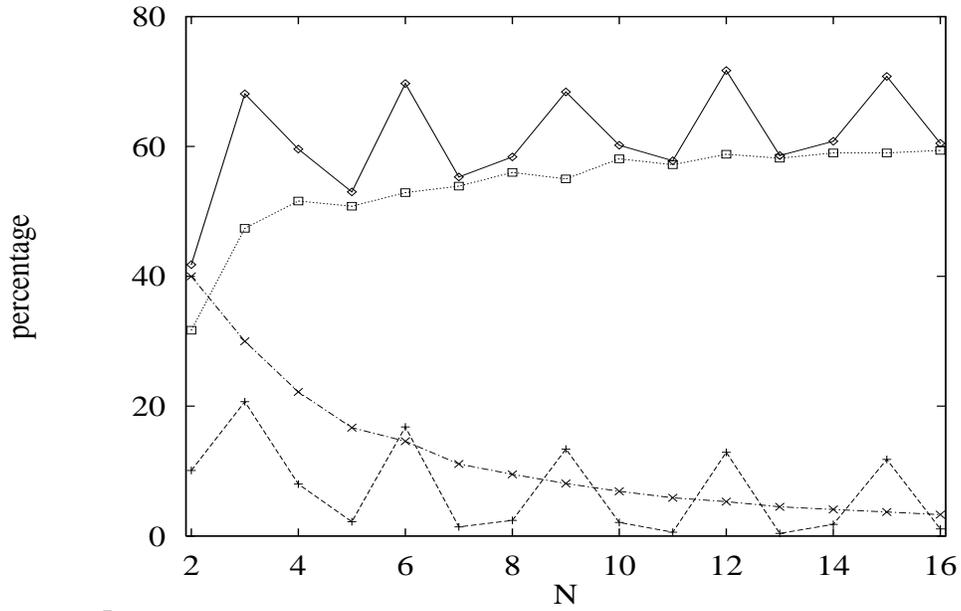,height=0.45\textwidth,width=0.75\textwidth} }}
\caption[]{Percentage of $L^P=0^+$ ground states as a function of 
the boson number $N$ for $H=H_2$ for which the energy ratio of 
Eq.~(\protect\ref{ratio}) is given by $0<R<1$ (dashed line), 
$R \geq 1$ (dotted line) and $R>0$ (solid line). The dashed-dotted 
line shows the percentage of $L^P=0^+$ states in the model space.}
\label{zero}
\end{figure}

We first analyze the dependence on the total number of bosons. 
Hereto we take the Hamiltonian $H_2$ of Eq.~(\ref{h2}) with random 
two-body matrix elements. In all calculations we make 1000 runs. 
For each set of randomly generated two-body matrix elements we 
calculate the entire energy spectrum and the $B(E2)$ values 
between the yrast states. In Fig.~\ref{zero} we show the 
percentage of $L^P=0^+$ ground states as a function of $N$ (solid 
line). For $N=2$ the Hamiltonian matrix is a real-symmetric random 
matrix. For each value of the angular momentum $L$ the ensemble 
corresponds to GOE, whose level distribution is a semicircle with 
radius $\sqrt{4dv^2}$ and width $\sqrt{(d+1)v^2}$ \cite{review}. 
In this case, the percentage of ground states for a given value of 
$L$ is determined by the dimension $d$ of the Hamiltonian matrix: 
$d=2$ for $L=0$, 2 and $d=1$ for $L=4$. 
For $3 \leq N \leq 16$ the situation is completely different. 
The ensemble is now TBRE. 
The dominant angular momentum of the ground state is determined 
by the shapes of the level distributions as a function of the 
angular momentum, in particular by the tails, {\it i.e.} the 
higher moments, of the distributions (all have the same centroid). 
The distribution whose tail extends 
furthest is the most likely to provide the ground state. For a 
semicircular (GOE) or a Gaussian distribution (TBRE in the nuclear 
shell model \cite{French,Bohigas,Gervois}) the shape is completely 
detemined by the width. In these two cases, the dominance of 
$L^P=0^+$ ground states can be correlated to the widths of the 
distributions \cite{BFP2,oax00}. However, for a system of 
interacting bosons the TBRE distribution of eigenvalues is neither 
semicircular (except for $N=k$) nor Gaussian \cite{Kota}. 
There is no relation between the width (first moment) and 
higher moments of the distribution, which determine  
the dominant angular momentum of the ground state. 
Table~\ref{sigma} shows that the width increases with angular 
momentum, whereas the most likely value of the ground state 
angular momentum is $L^P=0^+$. 
In fact, the probability that the ground state has a certain 
value of the angular momentum is not really 
fixed by the full distribution of eigenvalues, but rather by that 
of the lowest one. Work is in progress to elucidate the form of 
these distributions in a schematic exactly solvable model 
\cite{Kusnezov}. 

Despite the different shapes of the TBRE level distributions 
for fermions and bosons we find, just as in the fermion case 
\cite{JBD,CJ,JBDT,BFP2}, a dominance ($\sim 60 \%$) of $L^P=0^+$ 
ground states in the IBM with $3 \leq N \leq 16$. 
This fraction is large compared to the percentage of $L^P=0^+$ 
states in the model space (solid and dashed-dotted lines in 
Fig.~\ref{zero}). 
The oscillations with maxima at $N=3n$ (multiple of 3) are 
due to the `unphysical' region of parameter space for which the 
energy ratio 
\ba
R &=& \frac{E(4^+_1)-E(0^+_1)}{E(2^+_1)-E(0^+_1)} ~,
\label{ratio}
\ea
is less than 1 (dashed line) corresponding to a level sequence 
$0^+_1$, $4^+_1$, $2^+_1$, rather than $0^+_1$, $2^+_1$, $4^+_1$ 
for $R>1$ (dotted line). The enhancement for $N=3n$ can be 
attributed to the existence of a $0^+$ state in which all $d$ bosons 
are organized into $n_{\Delta}=N/3$ triplets. This state has the 
$U(5)$ quantum numbers $|N,n_d,v,n_{\Delta},L>=|N,N,0,N/3,0>$ and  
can become the ground state if the vibrational spectrum is 
turned `upside down'. 

\begin{figure}
\centerline{\hbox{
\psfig{figure=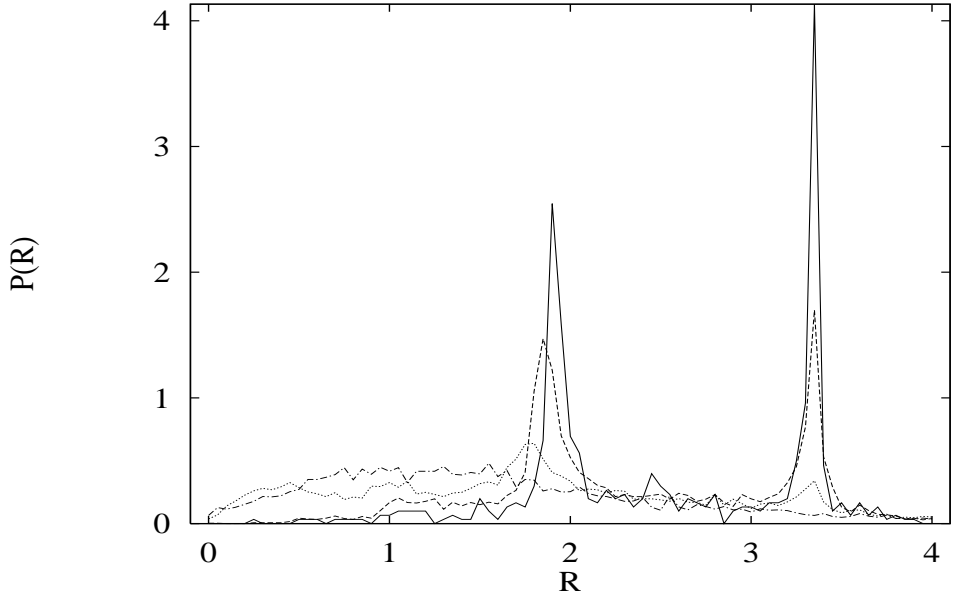,height=0.45\textwidth,width=0.75\textwidth}}}
\vspace{15pt}
\caption[H2]
{Probability distributions $P(R)$ of the energy ratio 
$R$ of Eq.~(\protect\ref{ratio}) with $\int P(R) dR = 1$ 
in the IBM with random two-body interactions 
for $N=3$ (dashed-dotted), 6 (dotted), 10 (dashed) and 16 (solid).} 
\label{pr2}
\end{figure}

For the cases with a $L^P=0^+$ ground state we present in 
Fig.~\ref{pr2} the probability distribution $P(R)$ of the energy 
ratio $R$ of Eq.~(\ref{ratio}). This energy ratio has very 
characteristic values for the harmonic vibrator and the rotor, 
$R=2$ and $R=10/3$, respectively. The Hamiltonian matrix of $H_2$ 
depends on 7 independent random two-body matrix elements. For small 
values of $N$ there is little correlation among the matrix 
elements of $H$, and as a consequence the probability distribution 
$P(R)$ shows little structure for $N=3$ (dashed-dotted curve). 
For increasing values of $N$ there is a correspondingly higher 
correlation between the different matrix 
elements of $H$, which results in the development of two peaks 
in $P(R)$. We first see the development of a maximum at 
$R \sim 1.9$ for $N=6$ (dotted curve), followed by another one 
at $R \sim 3.3$ for $N=10$ (dashed curve). For $N=16$ the  
probability distribution $P(R)$ has two very pronounced peaks, 
one at $R \sim 1.95$ and a narrower one at $R \sim 3.35$ (solid 
curve). These values correspond almost exactly to those for the 
harmonic vibrator and the rotor. 
The two maxima correspond 
to the two basic phases that characterize the collective region: 
a spherical one with $R \sim 2.0$ and a axially deformed one with 
$R \sim 3.3$. There is no peak for $\gamma$-unstable nuclei 
($SO(6)$ limit), since this requires that the matrix element of 
$\zeta_{2_{12}} \, \left[ (s^{\dagger} \times d^{\dagger})^{(2)} 
\cdot (\tilde{d} \times \tilde{d})^{(2)} + \mbox{h.c.} \right]$ 
vanishes identically, effectively corresponding to a zero-measure 
case for the random sample. Any other value of $\zeta_{2_{12}}$ 
($\neq 0$) gives rise to an axially symmetric rotor \cite{GK,DSI}. 

\begin{figure}
\centerline{\hbox{
\psfig{figure=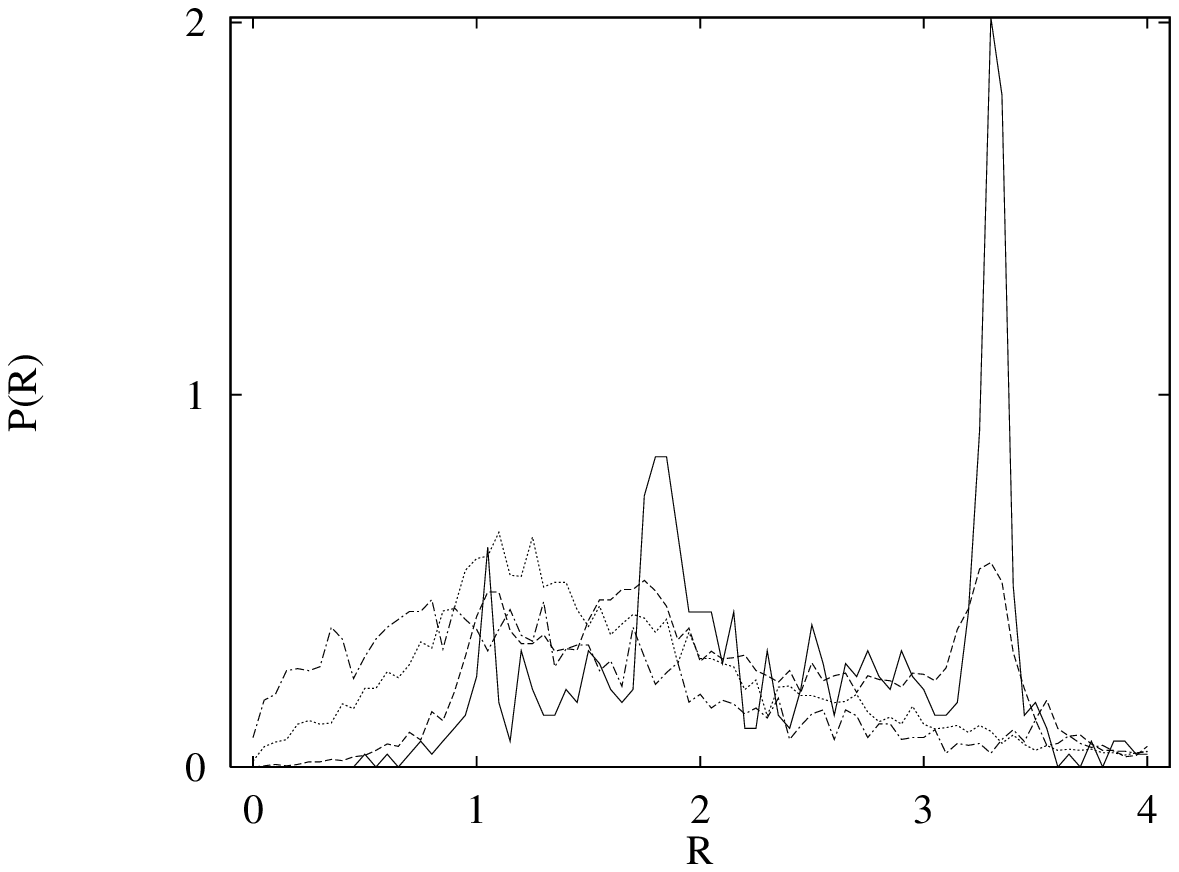,height=0.45\textwidth,width=0.75\textwidth}}}
\vspace{15pt}
\caption[H3]{As Fig.~\protect\ref{pr2}, but for 
random three-body interactions.} 
\label{pr3}
\end{figure}

In a second calculation we take the Hamiltonian $H_3$ of 
Eq.~(\ref{h3}) with random three-body interactions. Fig.~\ref{pr3} 
shows the same qualitative behavior as Fig.~\ref{pr2} although the 
peak structure is far less pronounced. For $N=16$ we see again two 
maxima at the vibrator and rotor values of the energy ratio $R$. 
The case of three-body interactions in the IBM is of special 
interest, since it can give rise to stable triaxial deformations 
\cite{Piet}, which are absent in the case of Hamiltonians including 
one- and two-body interactions only. We note, however, that in 
the neutron-proton version of the IBM, triaxial deformation can be 
obtained from Hamiltonians with one- and 
two-body interactions only \cite{DB}. Fig.~\ref{pr3} shows no 
clear sign of a `triaxial' peak ({\em e.g.} a triaxially deformed 
rotor with $\gamma=30^{\circ}$ has $R=8/3$), nor of a 
`$\gamma$-unstable' one with $R=5/2$. 

\begin{figure}
\centerline{\hbox{
\psfig{figure=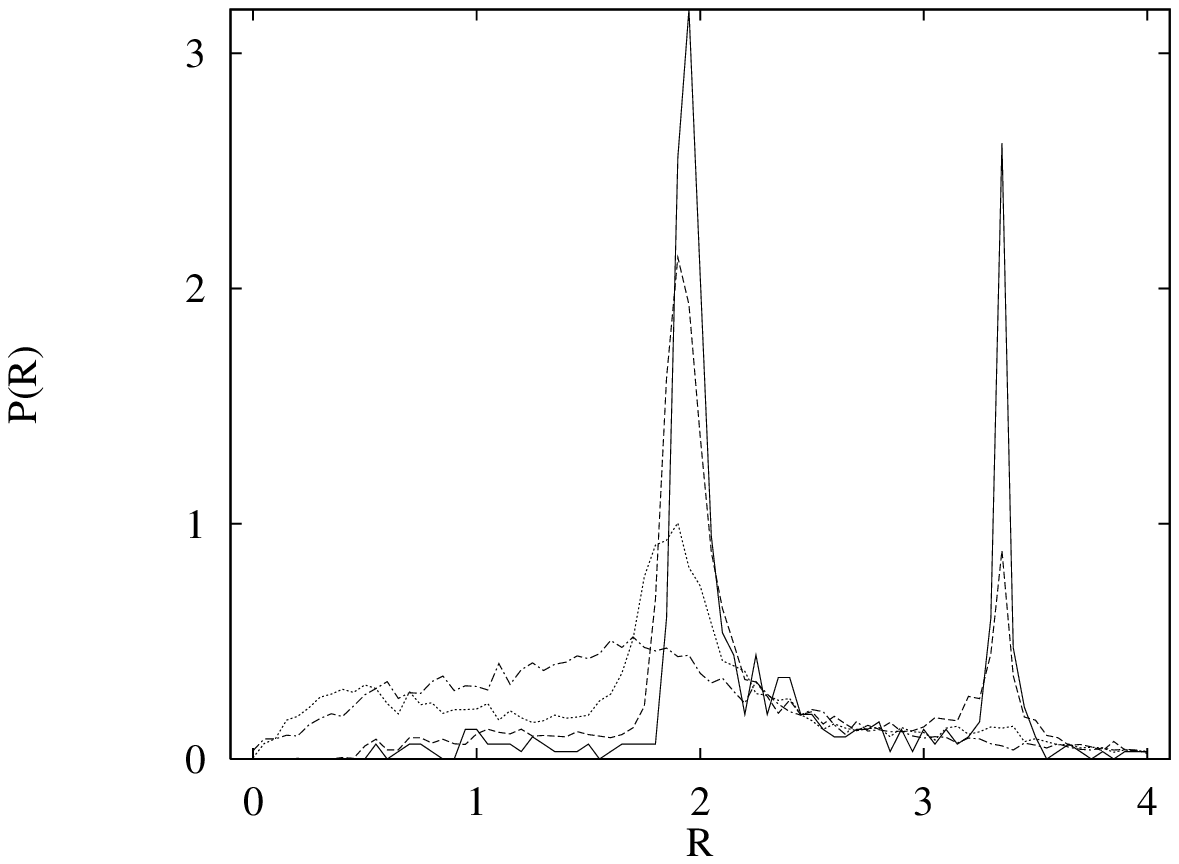,height=0.45\textwidth,width=0.75\textwidth}}}
\vspace{15pt}
\caption[H12]{As Fig.~\protect\ref{pr2}, but for 
random one- and two-body interactions.}
\label{pr12}
\end{figure}

\begin{figure}
\centerline{\hbox{
\psfig{figure=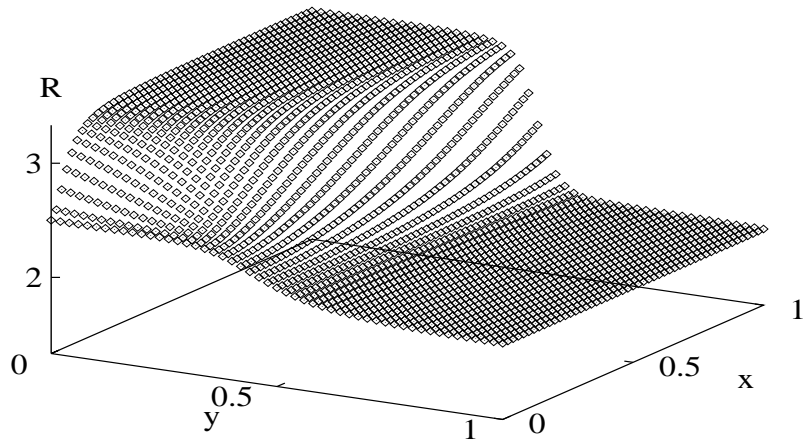,height=0.45\textwidth,width=0.75\textwidth} }}
\vspace{15pt}
\caption[]{The energy ratio $R$ of Eq.~(\protect\ref{ratio}) 
as a function of $x$ and $y$ in the consistent 
Q-formulation of the IBM.}
\label{cqf3}
\end{figure}

After these two model studies, we now turn to a more realistic 
case. It has been shown \cite{IBM} that the phenomelogy of 
low-lying collective states in nuclei is well described by an 
IBM Hamiltonian consisting of both one- and two-body interactions 
\ba
H_{12} &=& \frac{1}{N} \left[ H_1 + \frac{1}{N-1} H_2 \right] ~.
\label{h12}
\ea
In order to 
remove the $N$ dependence of the matrix elements of $k$-body 
interactions, we have scaled $H_k$ by $\prod_{i=1}^k (N+1-i)$. 
In Fig.~\ref{pr12} we show the corresponding probability 
distribution $P(R)$ of the energy ratio $R$ of Eq.~(\ref{ratio}) 
for different values of the number of bosons. The results are very 
similar to those of Fig.~\ref{pr2} which were obtained with pure 
two-body interactions. 
With increasing values of $N$ the many-body matrix elements 
of $H_{12}$ become increasingly correlated, which results in the 
development of two maxima in $P(R)$. The curve for $N=16$ is 
identical to the calculation discussed in \cite{BF}. 
The occurrence of two basic phases for the collective region is 
further exemplified in Fig.~\ref{cqf3} in which we plot the energy 
ratio $R$ for the consistent-Q formulation \cite{CQF} of the IBM 
\ba
H &=& \epsilon \, \hat n_d 
- \kappa \, \hat Q(\chi) \cdot \hat Q(\chi) ~, 
\nonumber\\
\hat Q_{\mu}(\chi) &=& ( s^{\dagger} \tilde{d} 
+ d^{\dagger} s)^{(2)}_{\mu} 
+ \chi \, (d^{\dagger} \tilde{d})^{(2)}_{\mu} ~, 
\label{hcqf}
\ea
with realistic values of the interactions, {\em i.e.} a positive $d$ 
boson energy ($\epsilon>0$) and an attractive quadrupole-quadrupole 
interaction ($\kappa>0$). The results in Fig.~\ref{cqf3} are 
plotted as a function of the scaled parameters $x=-2\chi/\sqrt{7}$ 
and $y=\epsilon/[\epsilon+4\kappa(N-1)]$, which have been used as 
control parameters in a study of phase transitions in the IBM 
\cite{DSI,IZC,Dimitri}). For $y=1$ we recover the vibrational or  
$SU(5)$ limit of the IBM, whereas for $y=0$ and $x=1$ 
one finds the rotational or $SU(3)$ limit, and for $y=0$ and 
$x=0$ the $\gamma$-unstable or $SO(6)$ limit. We clearly 
see two planes corresponding to $R \sim 2.0$ and $R \sim 3.3$ 
respectively, 
which are separated by a sharp transitional region, in agreement 
with the observation in \cite{IZC,Dimitri} that the collective 
region is characterized by two phases (spherical and deformed) 
connected by a sharp phase transition.  

\begin{figure}
\centerline{\hbox{
\psfig{figure=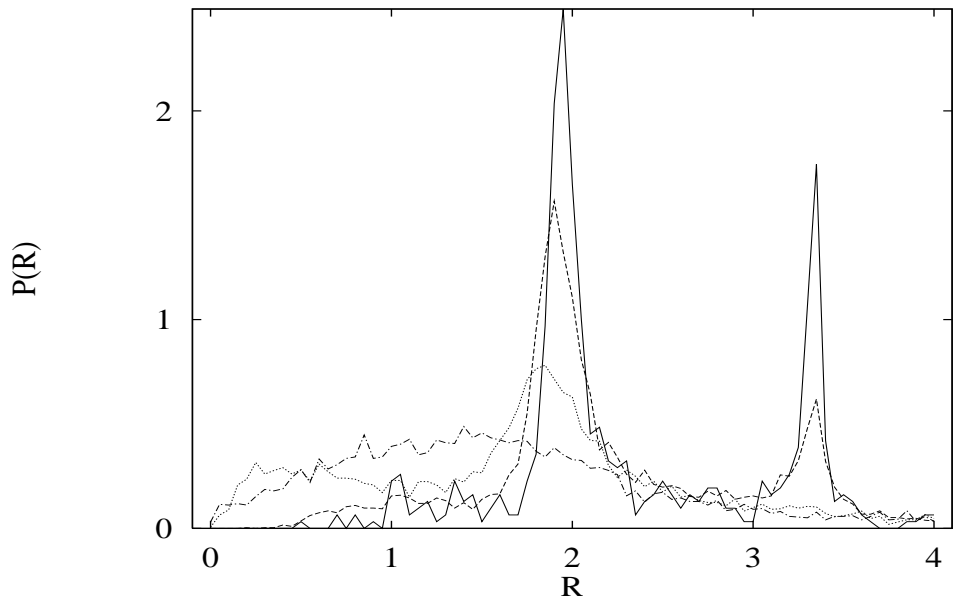,height=0.45\textwidth,width=0.75\textwidth}}}
\vspace{15pt}
\caption{As Fig.~\protect\ref{pr2}, but for 
random one-, two- and three-body interactions.} 
\label{pr123}
\end{figure}

In order to investigate the effect of higher order interactions 
we now add three-body interactions to the Hamiltonian 
\ba
H_{123} &=& \frac{1}{N} \left[ H_1 + \frac{1}{N-1} \left[ H_2  
+ \frac{1}{N-2} H_3 \right] \right] ~.
\label{h123}
\ea
In this case the Hamiltonian matrix depends on 26 independent 
random matrix elements (2 one-body, 7 two-body and 17 three-body). 
Therefore, for a fixed value of $N$ there is less correlation 
between the $N$-body matrix elements of $H_{123}$ than for $H_{12}$, 
which results in broader peaks in the probability 
distribution $P(R)$. A comparison of Figs.~\ref{pr12} 
and \ref{pr123} shows that the probability distribution $P(R)$ 
behaves in a very similar way, and that the addition of three-body 
interactions does not change the results in a significant way. 
When $N$ is sufficiently large compared to the maximum rank $k$ 
of the interactions (2 and 3, respectively) the results become 
independent of $k$. 

This result is qualitatively very similar to 
that of \cite{French}, in which the transition from a Gaussian to 
a semicircular level distribution was studied in the nuclear shell 
model for a fixed particle number $N=7$ with increasing values of 
the rank $2 \leq k \leq 7$. The characteristic features of the 
ensemble depend on the ratio of the number of particles and 
the rank of the interactions. For $N$ sufficiently large  
compared to $k$ there is a saturation, and the properties of the 
ensemble no longer depend on $k$. 

\section{Summary and conclusions}

In summary, we have studied global properties of low-lying 
collective levels using the interacting boson model with 
random interations. In particular, we addressed the dependence 
of the dominance of $L^P=0^+$ ground states and the occurrence 
of vibrational and rotational band structures on the boson number 
$N$ and the rank $k$ of the interactions.

Just as for the nuclear shell model it was 
found that despite the randomness of the interactions (both in 
size and sign) the ground state has $L^P=0^+$ in approximately 
60 $\%$ of the cases. The oscillation in the percentage of $L^P=0^+$ 
ground states with $N$ was shown to be entirely due to cases in 
which the level sequence is given by $0^+$, $4^+$, $2^+$ ($R<1$). 
For the cases with $R>1$ there is a very smooth dependence on $N$. 

The vibrational and rotational band structures appear gradually 
as $N/k$ increases. For $N \sim k$ there is little or no evidence 
for such bands. As $N$ grows we first see evidence for the 
development of vibrational structure, followed later by the 
appearance of rotational bands. If 
$N$ increases further these band structures become more and more 
pronounced. Essentially the same behavior is found for random two- 
and three-body interactions. 
In realistic applications to collective nuclei the IBM Hamiltonian 
consists of a combination of one- and two-body interactions. 
A study with random ensembles of one- and two-body interactions 
shows similar results to the case of pure two-body terms. 
The inclusion of random three-body 
interactions does not significantly change the basic features. 

In conclusion, we find that the dominance of $L^P=0^+$ ground states 
and the occurrence of vibrational and rotational features are 
independent of the boson number, as long as $N$ is sufficiently 
large compared with the maximum rank of the interactions. 
We can conclude that these features represent general and 
robust properties of the interacting boson model space, and are 
a consequence of the many-body dynamics, which enters via the 
reduction formulas for the $N$-body matrix elements of $k$-body 
interactions (angular momentum coupling, coefficients of fractional 
parentage, etc.). Since the structure of the model space is 
completely determined by the corresponding degrees of freedom, 
these results emphasize the importance of the selection of the 
relevant degrees of freedom. In this context, a relevant question 
is whether vibrational and rotational collective behavior can be 
directly observed in the shell model with random interactions if 
an appropriate truncation of the (shell model) Hilbert space is 
carried out. 

It is important to stress that these properties do not arise as 
an artefact of a particular model of nuclear structure. 
In empirical studies of the low-lying collective states of medium 
and heavy even-even nuclei very regular and robust features have 
been observed, such as the tripartite classification into seniority, 
anharmonic vibrator and rotor regimes \cite{Casten,Zamfir} and the 
systematics of excitation energy and $M1$ strength of the scissors 
mode \cite{tijeras}.  

Finally, we remark that the use of random interactions to study 
the generic behavior of low-lying states has also found useful 
applications in many-body quantum systems of a different nature, 
such a quantum dots or small metallic particles \cite{Carlos}. 

\section*{Acknowledgements}

It is a pleasure to thank Rick Casten, Jorge Flores, Franco Iachello 
Stuart Pittel and Thomas Seligman for illuminating discussions. 
This work was supported in part by DGAPA-UNAM under project 
IN101997 and by CONACyT under projects 32416-E and 32397-E.

\begin{table}
\centering
\caption[]{Percentage of ground states with angular momentum $L$
and corresponding widths of TBRE level distributions. 
The results are obtained for 1000 runs and $N=16$ bosons. 
The widths are divided by $N(N-1)$.} 
\label{sigma} 
\vspace{15pt} 
\begin{tabular}{rrrccrrrc}
%\hline
& & & & & & & & \\
$L$ & dim & TBRE & width & & 
$L$ & dim & TBRE & width \\
& & & & & & & & \\
\hline
& & & & & & & & \\
 0   &  30 &  60.5 $\%$ &   0.45 & \hspace{0.5cm} & 
17   &  23 &   0.0 $\%$ &   0.45 \\
 2   &  51 &  12.9 $\%$ &   0.44 & & 
18   &  31 &   0.7 $\%$ &   0.46 \\
 3   &  21 &   0.0 $\%$ &   0.43 & & 
19   &  16 &   0.0 $\%$ &   0.46 \\
 4   &  64 &   0.0 $\%$ &   0.43 & & 
20   &  23 &   1.3 $\%$ &   0.47 \\
 5   &  35 &   0.0 $\%$ &   0.42 & & 
21   &  11 &   0.0 $\%$ &   0.48 \\
 6   &  70 &   0.1 $\%$ &   0.43 & & 
22   &  16 &   0.8 $\%$ &   0.49 \\
 7   &  42 &   0.0 $\%$ &   0.42 & & 
23   &   7 &   0.0 $\%$ &   0.51 \\
 8   &  71 &   0.4 $\%$ &   0.43 & & 
24   &  11 &   0.9 $\%$ &   0.52 \\
 9   &  44 &   0.0 $\%$ &   0.42 & & 
25   &   4 &   0.0 $\%$ &   0.54 \\
10   &  67 &   0.1 $\%$ &   0.43 & & 
26   &   7 &   1.1 $\%$ &   0.56 \\
11   &  42 &   0.0 $\%$ &   0.42 & & 
27   &   2 &   0.0 $\%$ &   0.58 \\
12   &  60 &   0.2 $\%$ &   0.43 & & 
28   &   4 &   0.7 $\%$ &   0.60 \\
13   &  37 &   0.0 $\%$ &   0.43 & & 
29   &   1 &   0.0 $\%$ &   0.63 \\
14   &  51 &   0.3 $\%$ &   0.44 & & 
30   &   2 &   0.6 $\%$ &   0.64 \\
15   &  30 &   0.0 $\%$ &   0.44 & & 
32   &   1 &  18.6 $\%$ &   0.71 \\
16   &  41 &   0.8 $\%$ &   0.45 & & & & & \\
& & & & & & & & \\
%\hline
\end{tabular}
\end{table}

\end{document}